\documentstyle[11pt,newpasp,twoside,epsf]{article}
\begin{document}
\title{Oxygen line formation in 3D hydrodynamical model atmospheres
}
 \author{Martin Asplund}
\affil{Astronomiska Observatoriet, Box 515, SE-751 20 Uppsala,
Sweden 
%(martin@astro.uu.se)
}

\begin{abstract}
The formation of [O\,{\sc i}], O\,{\sc i} and OH lines in metal-poor stars
has been studied by means of 3D hydrodynamical model atmospheres. 
For O\,{\sc i} detailed 3D non-LTE calculations have been
performed. While the influence of 3D model atmospheres is minor
for [O\,{\sc i}] and O\,{\sc i} lines, the very low temperatures 
encountered at low metallicities have a drastic impact on OH. 
As a result, the derived O abundances are found
to be systematically overestimated in 1D analyses,
casting doubts on the recent claims for a monotonic increase 
in [O/Fe] towards lower metallicities.
\footnote{The full version of the present talk with additional figures can be found at http://www.astro.uu.se/$\sim$martin/talks/Manchester00}

\end{abstract}

\section{Introduction}

Besides being relatively common, oxygen has attracted a great
deal of attention due to its special role in galactic and stellar evolution.
The ratio [O/Fe] at different cosmic epochs 
give important insight to the formation and evolution of the Galaxy, as
well as constraining the physics of supernovae.
Oxygen can also significantly influence stellar evolution 
and is crucial for the production of the light elements Li, Be and B 
through cosmic ray spallation.  
Following the first indication of an oxygen over-abundance relative to iron 
%([O/Fe]~$>0$) 
in metal-poor stars, a great number of
studies have been devoted to quantifying this enhancement
but with little concordance. 
%Although all agree on its existence, the amount of 
%the over-abundance is hotly contested. 
%Various oxygen diagnostics in different types of stars
%have been applied with disparate results. 
The forbidden [O\,{\sc i}] line
at 630.0\,nm in metal-poor giants suggest a nearly flat
plateau at [O/Fe]\,$\sim 0.5$ for [Fe/H]\,$\la -1.0$ 
while the O\,{\sc i} triplet at 777\,nm in dwarfs
tend to imply systematically higher values, often with a monotonic
increase towards lower metallicities.
Recently, Israelian et al. (1998) and Boesgaard et al. (1999) have
analysed the UV OH 
%(A$^2\Sigma$ - X$^2\Pi$) 
lines down to [Fe/H]~$\simeq -3.3$ and have found a linear trend in 
[O/Fe] vs [Fe/H] with
a slope of about -0.4, in stark conflict with the [O\,{\sc i}] results.
All oxygen criteria have their pros and cons, which influence the
conclusions. Of particular interest is the possible systematic errors 
introduced by the inherent assumptions behind the 1D model atmospheres 
employed in the analyses of the various O diagnostics.
%(plane-parallel, homogeneous, hydrostatic, LTE stratification with convection 
%treated solely through the simple-minded mixing length ``theory'').
Significant progress in this respect has recently been accomplished
with the construction of 3D stellar convection simulations 
%for both solar-type and metal-poor stars 
(e.g. Stein \& Nordlund 1998; Asplund et al. 1999, 2000a). 
The aim of the present investigation is to study
the impact of such improved models on oxygen line formation. 
%Could possibly the enigmatic oxygen dilemma
%which is the topic of this Joint Discussion finally be resolved by 
%the new generation of 3D, time-dependent hydrodynamical model atmospheres?

\section{3D hydrodynamical model atmospheres and line formation}

Realistic {\em ab-initio} 3D, time-dependent simulations of stellar surface
convection form the foundation
for the present study. The same incompressible radiative hydrodynamical code
which previously has been applied successfully to studies of solar
and stellar granulation 
(e.g. Stein \& Nordlund 1998; Asplund et al. 1999, 2000a)
%; Asplund \& Carlsson 2000)
have here been used to construct sequences of 3D model atmospheres. 
Special care have been taken to include the most appropriate input physics
in terms of equation-of-state, opacities and line-blanketing. 
%These convection simulations have here been employed as 3D model atmospheres
%for calculations of the line formation process of 
%O\,{\sc i}, [O\,{\sc i}] and OH lines. For the latter two types of 
%transitions local thermodynamic equilibrium (LTE) has been assumed while for
%the former detailed 3D non-LTE calculations have been performed.
Further details on the numerical procedures 
may be found in the above-mentioned references. 

\begin{figure}
\label{f:[OI]}
\plottwo{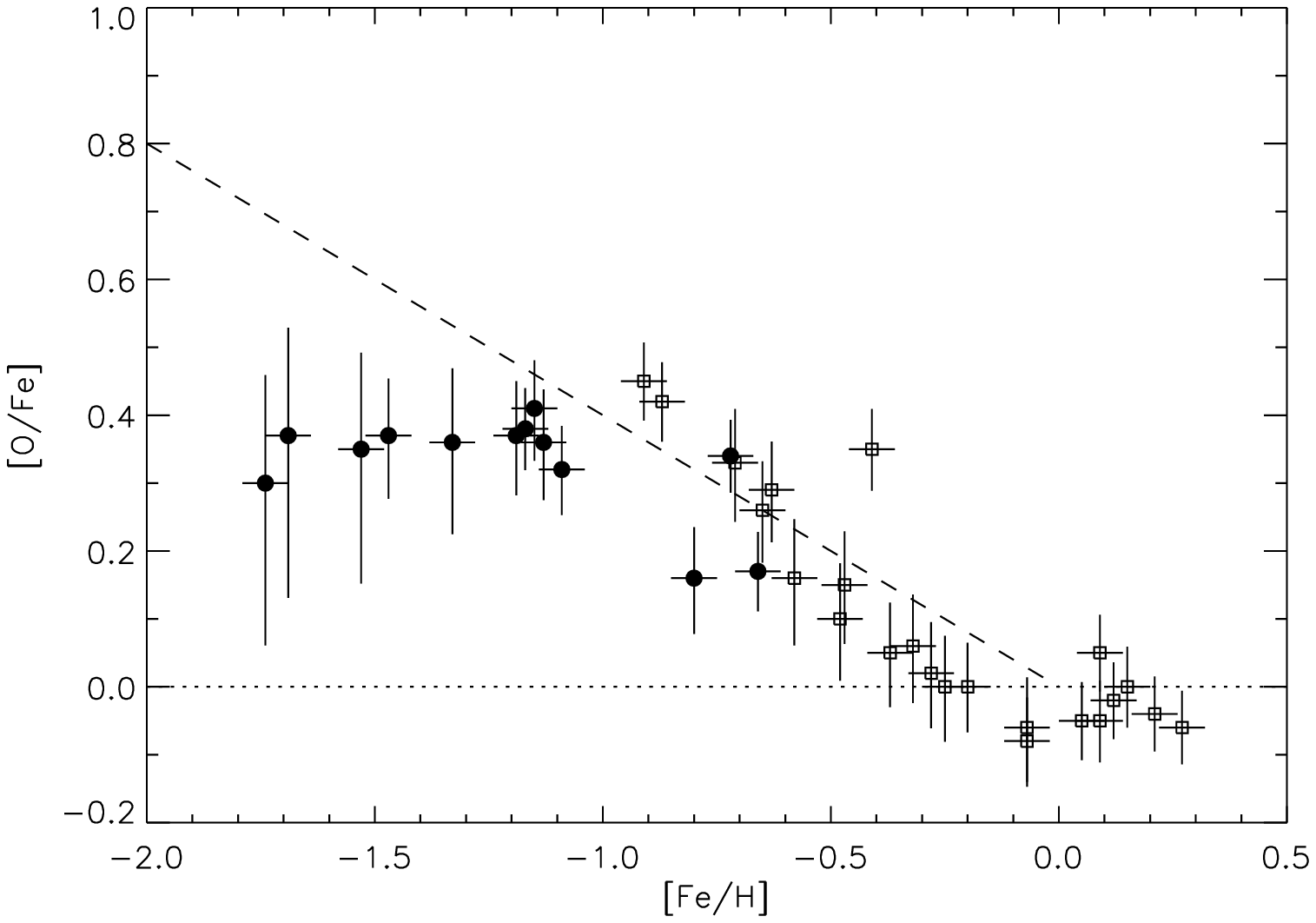}{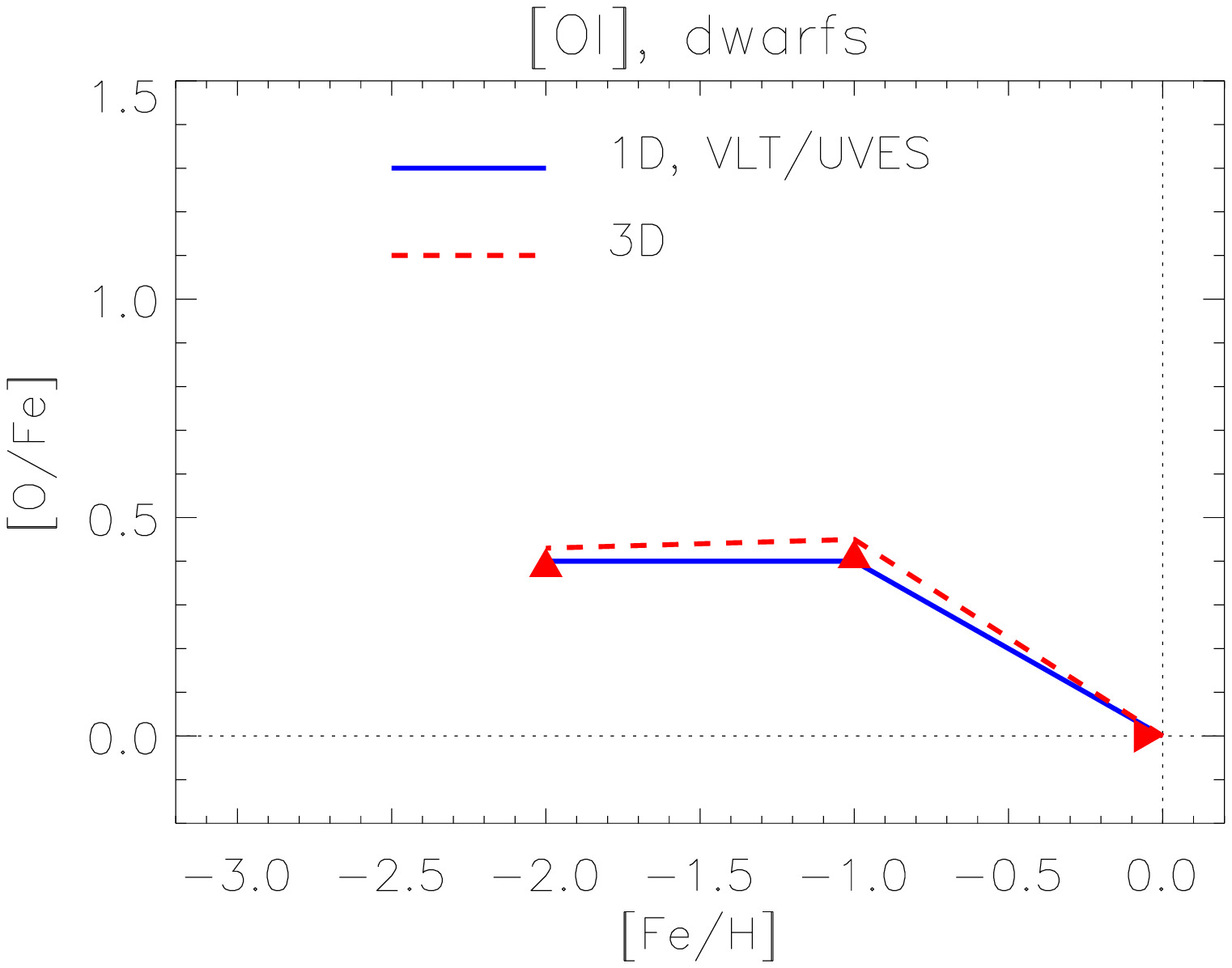}
\caption{{\em Left panel:} [O/Fe] ratios determined from [O\,{\sc i}]
and Fe\,{\sc ii} lines from very high $S/N$ VLT spectra
(Nissen et al., in preparation). 
{\em Right panel:} The (minor) influence of 3D models on
the [O\,{\sc i}] line}
\end{figure}

The 3D model atmospheres differ from classical 1D models both in their
average photospheric structures and through the presence of
temperature inhomogeneities and convective
flows, which all influence the spectral line formation
(e.g. Asplund et al. 2000a,b). 
In particular for metal-poor stars, the 3D simulations differ radically
from 1D model atmospheres (Asplund et al. 1999).
%While the temperature remains close to the radiative
%equilibrium value at solar metallicities and 
%mild metal-deficiencies ([Fe/H]$\ga -1.0$),
%the temperature in the outer layers depart significantly from it at 
%lower metallicities.  
Due to the dominance of adiabatic cooling over radiative heating in
the optically thin layers, very low temperatures 
are encountered in the metal-poor simulations compared with
the 1D radiative equilibrium expectations.
As a result, the 3D models have a profound impact on the line
formation of in particular low-excitation lines of minority species and
molecular lines, such as Li\,{\sc i} and OH. 

\section{Impact on O abundances in metal-poor stars}

%\subsection{[O\,{\sc i}]}

Since essentially all of O is in the ground state of O\,{\sc i}
and the [O\,{\sc i}] line
formation is very well described by LTE, the influence of 3D models
on the [O\,{\sc i}] line is minor. As a consequence, the [O/Fe] plateau
appears robust, provided high $S/N$ spectra and Fe\,{\sc ii} lines
are employed (Fig. 1 and Nissen et al., in preparation).

%\subsection{O\,{\sc i}}

Kiselman (these proceedings) has given convincing arguments that 
the O\,{\sc i} triplet suffers from departures from LTE. 
The O\,{\sc i} line formation
has therefore been investigated through detailed 3D non-LTE calculations
for the Sun and HD\,140283 ([Fe/H]~$=-2.5$)
using {\sc multi3d} (Botnen 1997) and an extensive 23-level O model atom.
Very similar non-LTE abundance corrections to the 1D non-LTE results
(Kiselman 1993) are found, typically 0.2\,dex with only a minor
metallicity dependence. Thus, 3D models will unlikely provide a mean
to bring down the [O/Fe] trend to plateau-like values (Fig. 2). 
%Instead different $T_{\rm eff}$-calibrations may be the solution
%(Gustafsson, these proceedings).

%\subsection{OH}

The low atmospheric temperatures in the metal-poor 3D models become
very apparent in the OH line strengths. Since molecular formation
is extremely temperature sensitive ($N_{\rm OH} \propto T^{-12...-14}$),
1D model atmospheres are bound to severely
underestimate the OH line strengths, and consequently overestimate
the O abundances at low metallicities (Fig. 2). 
The here presented 3D LTE OH analysis therefore casts 
doubts on recent claims of large
O over-abundances in metal-poor stars.
The OH line formation is studied further in 
Asplund \& Garc\'{\i}a P{\'e}rez (2000).

\begin{figure}
\label{f:OI+OH}
\plottwo{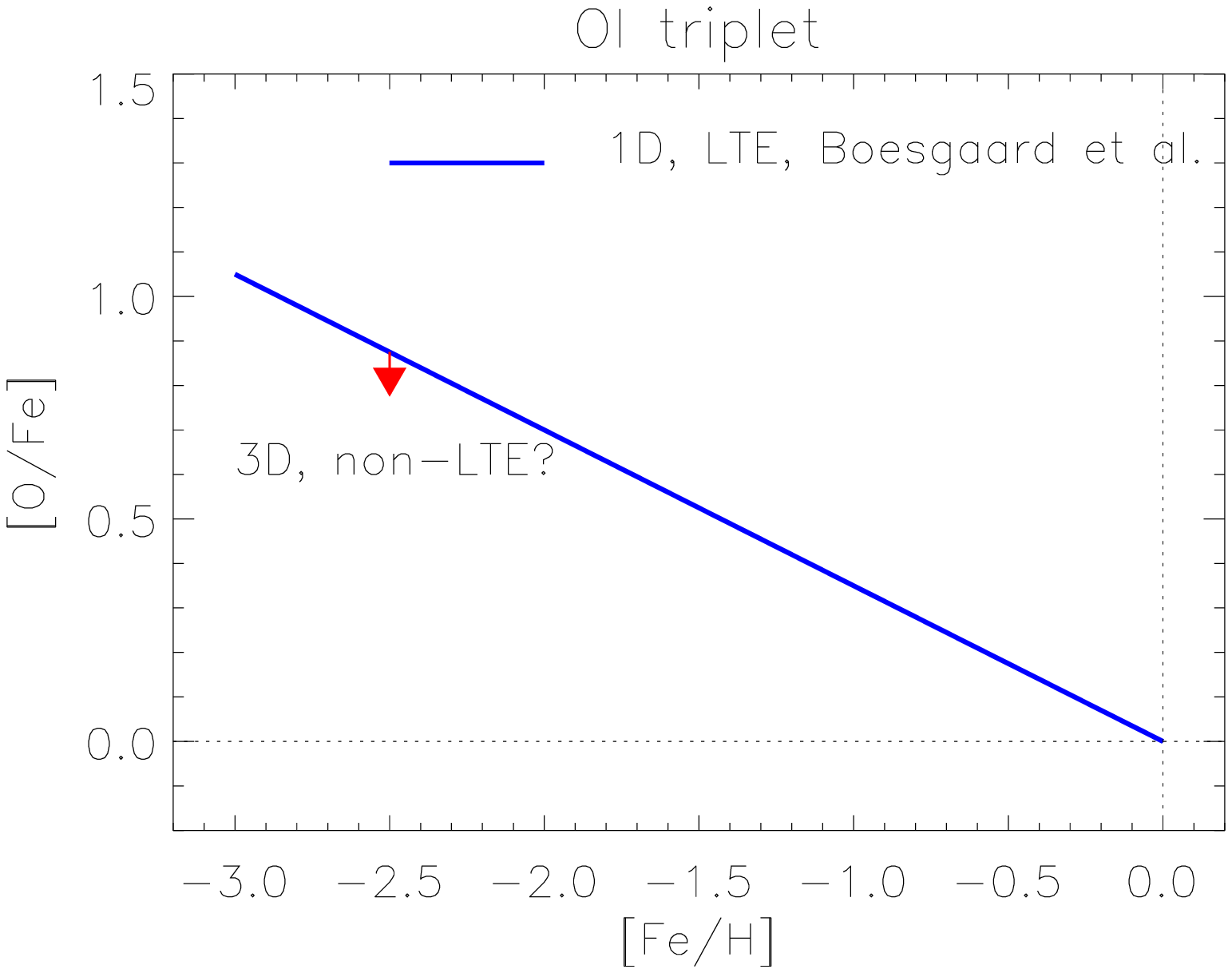}{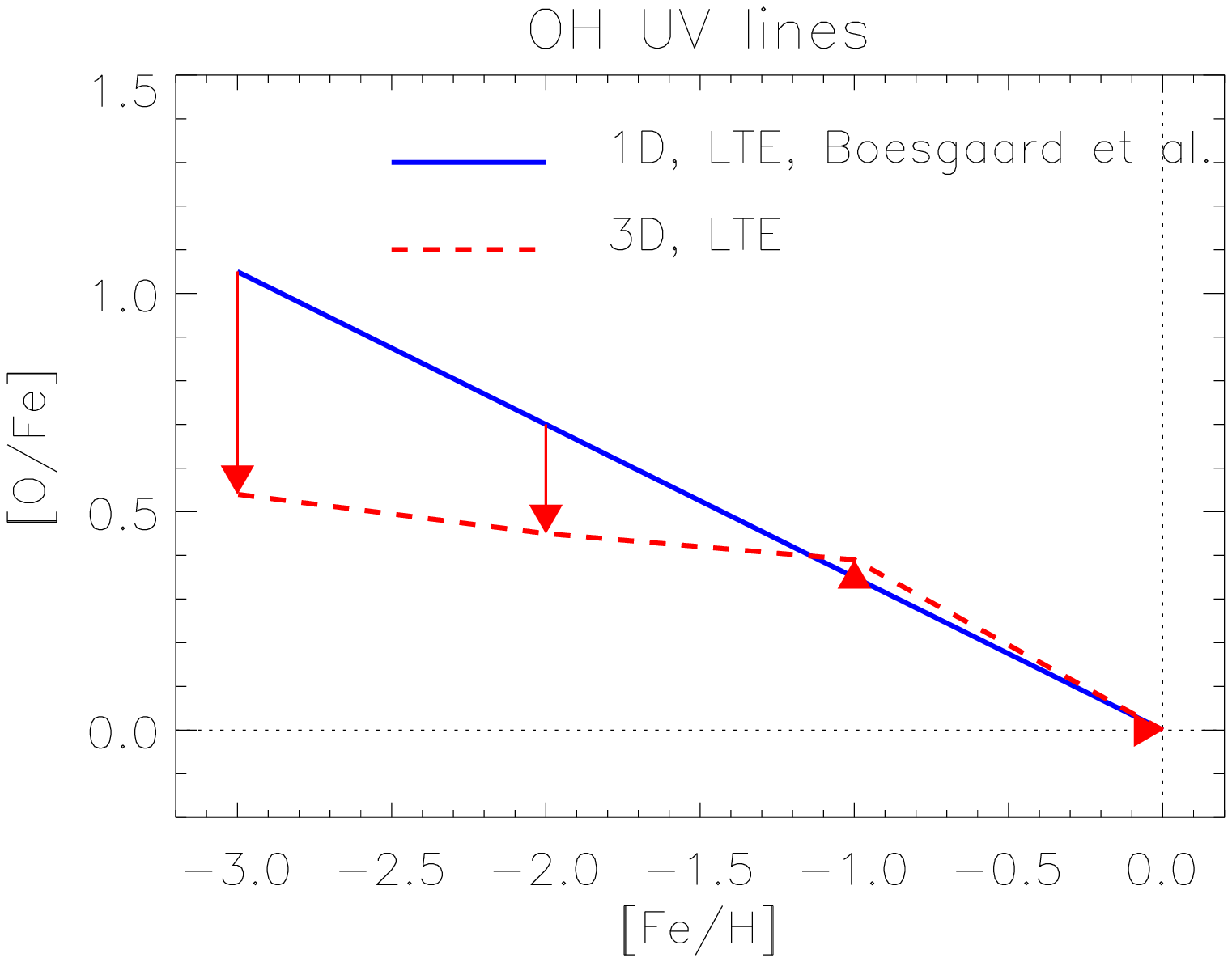}
\caption{ 
The influence of 3D models on the O\,{\sc i} triplet {\em (Left panel)}
and the UV OH lines {\em (Right panel)}}
\end{figure}

\section{Conclusions}

The new generation of 3D hydrodynamical model atmospheres has 
a dramatic impact on in particular the OH line formation in
metal-poor stars. Caution must, however, be
exercised in interpreting these 3D findings in terms of galactic
cosmic evolution until detailed 3D non-LTE calculations for the
OH molecular formation and line formation have been performed.

%%%%%%%%%%%%%%%%%%%%%%%%%%%%%%%%%%%%%%%%%%%%%%%%%%%%%%%%%%%%%%%
\acknowledgments
The author greatly appreciates very stimulating collaborations with 
among others M. Carlsson, A.E. Garc\'{\i}a P{\'e}rez, 
D. Kiselman, D.L. Lambert, P.E. Nissen, \AA. Nordlund and F. Primas.

%%%%%%%%%%%%%%%%%%%%%%%%%%%%%%%%%%%%%%%%%%%%%%%%%%%%%%%%%%%%%%%

\end{document}